\documentclass[12pt]{article}
\usepackage{amssymb,amsfonts}

\input{psfig}
\newcommand{\nc}{\newcommand}

\newcommand{\nn}{\nonumber}

\nc{\beq}{\begin{equation}}
\nc{\eeq}{\end{equation}}
\nc{\beqa}{\begin{eqnarray}}
\nc{\eeqa}{\end{eqnarray}}
\nc{\lsim}{\begin{array}{c}\,\sim\vspace{-21pt}\\< \end{array}}
\nc{\gsim}{\begin{array}{c}\sim\vspace{-21pt}\\> \end{array}}
\nc{\eps}{\epsilon}
\nc{\s}{\sigma}
\nc{\veps}{\varepsilon}
\nc{\no}{\noindent}
\nc{\D}{\Delta}
\nc{\al}{\alpha}
\nc{\be}{\beta}
\nc{\ga}{\gamma}
\nc{\de}{\delta}
\nc{\dmi}{{1\over 2}}
\begin{document}
\begin{flushright}
{PAR-LPTHE 98/20}  
\end{flushright}

\begin{center}
\vspace{1cm}
{\large \bf 
A Renormalization Group Study of Asymetrically Coupled Minimal Models}
 
\vspace{1cm}
 
  {\bf  M.-A. Lewis\footnote{lewism@lpthe.jussieu.fr}, P. Simon\footnote{simon@
lpthe.jussieu.fr}}
 
\vspace{.5 cm}
 
 {\it   Laboratoire de Physique Th\'eorique et Hautes Energies  }\footnote{ 
 Unit\'e associ\'ee au CNRS URA 280}\\
 {\it  Universit\'es Pierre et Marie Curie Paris VI et Denis Diderot Paris 
VII}\\
{\it  2 pl. Jussieu, 75251 Paris cedex 05 }\\
{\it FRANCE}
  \vspace{1cm}   
\end{center}
\vspace{1cm}

%
\begin{abstract} We investigate the renormalization group flows and fixed point structure of many coupled
  minimal models.  The models are coupled two by two by energy-energy 
couplings.  We take the general approach where the bare couplings are all taken
 to be independent.  New fixed points are found for $N$ models ($N\ge 3$).  At 
these fixed points, the coupling constants all have the same magnitude, but some are positive while others are negative.  By analogy with spin lattices, these can be interpreted as non-frustrated configurations with a maximal number of antiferromagnetic links.  The stability of the different fixed points is studied. We compute the critical exponents and spin-spin correlation functions between different models.  Our classification is shown to be complete.       
\end{abstract}
%
%

\newpage

A better understanding of coupled conformal field theories is primordial for the study of
various problems in condensed matter theory as in the so-called spin-ladders or in disordered spin systems.  Moreover, it could help to learn something about three dimensional theories. A promising approach to perturbed conformal field theory is the theory of
 integrable perturbations \cite{vays,LLM,GZ}.  Here, we choose the more modest approach of perturbative conformal field theory introduced in the pionnering work of Zamolodchikov \cite{zamolod}. Namely, we study the fixed point 
structure of the renormalization group equations for several coupled identical minimal models.  The couplings are initially taken to be independent, so that we have
 one dynamical variable for each pair of models.  The RG equations are computed
 up the third power in the coupling constant by using the Coulomb gas representation of minimal conformal field theories.  These techniques have already been used in order to study the effects of a weak disorder on second order phase transitions \cite{Lu,DPP}.   In this picture, the minimal models are parametrized by $\epsilon$, which is related to their central charge deviation from the Ising model value $\frac{1
}{2}$.  
 
We present results for $N$ coupled models, $N>3$.  For all of these models, we find symmetric fixed points (all the couplings are equal in magnitude and sign), and new ``mixed'' fixed points, that is, points where all couplings have the same magnitude, but different signs.  The number 
and arrangement of the links is shown to be equivalent to a graph coloring problem.  The solution to this problem is presented, and the resulting configurations are shown to be, by analogy to spin lattices, totally non-frustrated with a maximal number of antiferromagnetic links.  The stability of the new fixed points is then studied, and we find that these are tricritical points.  A study of the RG flows and the physical nature of the fixed points enables us to dress an accurate picture of the different phases. 
We then consider some of the critical exponents and the spin-spin correlation functions between points from different models. 
We end by looking at the physical significance of the new fixed points and by some comments on possible numerical verification of these results.       
 
\vskip 0.5cm
 
We consider $N$ coupled minimal models with energy-energy couplings:
\begin{equation}
H_N = \sum_{i=1}^N H_0^{(i)} - \sum_{i\ne j}^N g_{ij} \int d^2x\,\varepsilon_i(x)\varepsilon_j(x),
\end{equation}
$H_0$ being the energy of a single unperturbed minimal model.  The partition function is given by
\begin{equation}
Z_N = \left[\prod_{i=1}^N \mbox{Tr}_i\right]\, e^{(\sum_{i=1}^N H_0^{(i)}) + H_I}, 
\end{equation} 
and therefore can be seen as the perturbation of a conformal field theory of $N$ decoupled minimal models by the interaction term
\begin{equation}
H_I = \sum_{i\ne j} g_{ij} \int d^2x \,\varepsilon_i(x)\varepsilon_j(x).
\end{equation}
We will study the renormalization of the coupling constants.  Assuming all couplings to be independent is not only the most general approach, but also the only reasonable one. 

For example, if we had begun with a chain of critical models coupled only to their nearest neighbours,  first loop computations would show that the chain develops next nearest neighbour couplings.  Higher order computations eventually couple every model to all others, in such a way that there are no perturbative fixed points compatible with the chain geometry.

\vspace{.25 cm}
 
\noindent{\bf Renormalization group equations}

Let us first begin with the case $N=3$ depicted in Figure 1.\\  
To compute the renormalization of the coupling constants, we use the Coulomb gas representation 
of minimal models \cite{DF1}.  In this picture, the central charge of the model is characterized by the parameter $\alpha_+^2 \equiv \frac{4}{3} + \epsilon$
\begin{equation}
c=1-24\alpha_0^2 \qquad \alpha_\pm = \alpha_0 \pm \sqrt{\alpha_0^2+1}.
\end{equation}
The parametrization of $\alpha_+$ is chosen is such a way that, for $\epsilon$ equal to zero, we
 recover the Ising model.  For instance, the three-states Potts model has $\epsilon=-\frac{2}{15
}$. 
 
The renormalization of $g_{ij}$ is computed by expanding the interaction term, and then summing 
up the diagrams that contribute to $\varepsilon_i(x)\varepsilon_j(x)$.  The calculation techniques
were used to study disordered systems (see for example \cite{DPP,pujol}), and were found to be in good agreement with numerical simulations (at least for the three-states Potts model).  Summing contributions up to two loops, and rewriting bare quantities in terms 
of renormalized ones, we get the following equations for three coupled models:
\begin{eqnarray}
\label{beta}
\beta(g_{12}(r)) \equiv r\frac{\mbox{d}g_{12}}{\mbox{d}r} &=&\epsilon g_{12} +  g_{13}g_{23} -\dmi g_{12}\left(g_{13}^2+g_{23}^2\right) + {\cal O}(g^4)  \nn\\
\beta(g_{13}(r)) \equiv r\frac{\mbox{d}g_{13}}{\mbox{d}r} &=& \epsilon g_{13} + g_{12}g_{23} - \dmi g_{13}\left(g_{12}^2+g_{23}^2\right)+ {\cal O}(g^4) \nn\\
\beta(g_{23}(r)) \equiv r\frac{\mbox{d}g_{23}}{\mbox{d}r} &=& \epsilon g_{23} +  g_{12}g_{13} -\dmi g_
{23}\left(g_{12}^2+g_{13}^2\right)+ {\cal O}(g^4) 
\end{eqnarray}
Notice that we redefined $(-3\eps\to \eps)$ and $(4\pi g_{ij}\to g_{ij})$ in order to lighten notations. 
We are looking for fixed points of the renormalization group, that is, set of couplings such that $\beta(g_{ij})
=0$ for all $i,j$.  This condition can be rewritten in a more convenient form as
\begin{eqnarray} 
g_{12} \left(-\epsilon + \dmi (g_{13}^2+g_{23}^2)\right) &=& g_{13} g_{23} \nonumber\\
g_{13} \left(-\epsilon + \dmi (g_{12}^2+g_{23}^2)\right) &=&  g_{12} g_{23} \nonumber\\
g_{23} \left(-\epsilon + \dmi (g_{12}^2+g_{13}^2)\right) &=&  g_{12} g_{13},\label{fp}
\end{eqnarray}
Solutions are then easily obtained.  Besides the trivial $g^*_{ij}=0$ solution (stars denoting fixed point values), there is the symmetric solution:
\begin{equation}
g^*_{12}=g^*_{13}=g^*_{23}=  (-\epsilon + \epsilon^2) + {\cal O}(\epsilon^3),
\end{equation}
which corresponds to the fixed point of $N$ symmetrically coupled minimal models \cite{pujol}. To see it, we project our space of coupling constants on the direction $g_{ij}=g$, and the beta functions (\ref{beta}) reduce to 
\begin{equation}
r\frac{\mbox{d}g}{\mbox{d}r}=\eps g+g^2-g^3+ {\cal O}(g^4)
\end{equation}
which has only one non-trivial fixed point $g^*=  (-\epsilon + \epsilon^2) + {\cal O}(\epsilon^3)$.

The system (\ref{fp}) also has an ``anisotropic'' or mixed solution given by
\begin{equation}
g_{13}= - \gamma_3 \qquad g_{13}=g_{23}=+ \gamma_3
\end{equation}
where
\begin{equation}
\gamma_3 = \epsilon -  \epsilon^2 + {\cal O}(\epsilon^3).
\end{equation} 

These are the only perturbative solutions (with no finite part for $\epsilon=0$) modulo the permutation of models. We will generalize this result to the case of $N$ coupled minimal models.  These calculations generalize easily to three different minimal models (parametrized by $\eps_1,\eps_2\eps_3$) with similar conclusions. The  involved integrals have been performed in \cite{simon}.  Evidently, other solutions for $N$ models can be generated from solutions for a lower number of models by direct sum.  

Let us first comment the particular case
$\eps=0$, which corresponds to three coupled critical Ising models.
If we consider three ``isotropically'' coupled Ising models, this system is described by a $3$-colored Gross-Neveu model whose infrared behavior depends of the sign of the initial coupling constant (asymptotically free theory or massive regime) \cite{fradkin}. When introducing anisotropy, the system (\ref{fp}) has a non-trivial fixed point (up to permutations) solution  $g^*_{12}=g^*_{23}=0$ which can be interpreted as one Ising model plus one Ashkin-Teller model. This fixed point can be reached only if $g_{12}^0<0,g_{23}^0<0$.  The $O(3)$ symmetry of the $3$-colored Gross-Neveu model is then broken to $SU(2)\times U(1)$.

We now return to the more general case $\eps\ne 0$ and study the stability of the new fixed points.
It can be determined by looking at infinitesimal variations in the vinicity of the fixed points. If we take 
\begin{equation}
g_{12} = \gamma + \delta g_{12}\qquad g_{12} = \gamma + \delta g_{12}  \qquad g_{13} = -\gamma + \delta g_{13},
\end{equation}
then from the RG equations (\ref{beta}), we have ($\vec g \equiv (g_{12},g_{23},g_{13})$)
\begin{equation}
\delta\dot{\vec g} = A \delta\vec g
\end{equation}
where (it is sufficient to retain only first order in $\eps$)
\begin{equation}
A=\left[ \begin{array}{ccc}
        \eps & -\eps  & \eps \\
        -\eps & \eps  & \eps \\ 
        \eps & \eps  & \eps  
\end{array}\right]
\end{equation}
Since $\epsilon\ge 0$ (we redefined $(\epsilon \rightarrow -3\epsilon)$ earlier), the matrix $A$ has two positive and one negative eigenvalues.  The mixed fixed points are thus
tricritical.  The stable direction is given by
\begin{equation}
\delta g_{stab} = \frac{1}{\sqrt{3}} \left(g_{12} + g_{23} - g_{13}\right).
\end{equation}

Calculations are easily generalized for an higher number of coupled models.  One still finds both isotropic and anisotropic solutions.  If one defines the matrix $g$ whose elements are $g_{ij}$, then the beta functions for general $N$ are
 
\begin{equation}
\label{betan}
\beta(g_{ij}) = \epsilon g_{ij} +  (g^2)_{ij} - \dmi g_{ij}((g^2)_{ii} + (g^2)_{jj} - 2(g_{ij})^2)
\end{equation}
 
The condition $\beta(g_{ij})=0$ again leads to the symmetric solution. 
\begin{equation}
g^*_{ij}= \frac{1}{(N-2)}\epsilon - \frac{1}{(N-2)^2} \epsilon^2 + {\cal O} (\epsilon^3),
\end{equation}
and to anisotropic configurations, with $g^*_{ij}= \pm \ga_N$, whose exact arrangement we will discuss below.  Evidently, other solutions for $N$ models can be generated from solutions for a lower number of models by direct sum (a solution for N coupled models can be constructed with solutions for $M$ and $N-M$ models ($M<N$)).  The value of the coupling $\ga_N$ is given by
\begin{equation}
\ga_N=\frac{1}{(N-2)}\epsilon - \frac{1}{(N-2)^2} \epsilon^2 + {\cal O} (\epsilon^3).
\end{equation} 

We can thus write all couplings as $g_{ij}^*=\tau_{ij}\ga_N$ with $\tau_{ij}=\pm 1$. At this point, we can make an interesting analogy with lattice  spin models. If $\tau_{ij}<0$, we can assume the coupling $g_{ij}$ to be ferromagnetic (F) else antiferromagnetic (AF).  If all bonds are ferromagnetic, $\tau_{ij}=-1$.  
The existence of ``mixed'' solutions, that is solution with both F and AF couplings, 
was explicitely shown for the case $N=3$ in a previous section. The anisotropic solution has two AF bonds and one F bond. Note that the product of the three coupling constants is also negative. By analogy to spin lattice systems (if we replace for example minimal models by genuine Ising spins), it corresponds to a non-frustrated solution. Note that for Ising spins, this solution is equivalent to the symetric one. Finding the general configuration of ferromagnetic and antiferromagnetic bonds for $N$ models is not obvious. However, by looking carefully at the renormalization equations, one can convince himself that solutions with couplings of
 different signs are possible if and only if every triplet of points $i,j,k$ is such that
\begin{equation}
\tau_{ij}\tau_{jk}\tau_{ik} = -1.
\end{equation}
which correspond to non-frustrated configurations.  This is surely the case for the purely ferromagnetic solutions.  To find other configurations, the problem can be reformulated in
 a more convenient way: 

{\em Consider a system of $N$ points.  Points are linked two by two in such a way that every triplet of points contains one or three links.  What are the possible configurations?  Are they unique (up to permutations of models)?}  

This graph theory problem is identical to our RG fixed point condition if the presence of a link is the analogue of a ferromagnetic bond and the absence of link the 
analogue of an antiferromagnetic one.  Besides the fully connected solution, there is a unique solution (up to permutation of models) such that every triplet of points is connected by one or three bonds.  The total number of bonds, $N_b$, is given by
\begin{equation}
N_b  =\left\{\begin{array}{ll} 
p^2~, & N=2p \\
p^2+p~, & N=2p+1.
\end{array}\right.
\end{equation}
The solution is constructed in the following way.  {\em Label sites from $1$ to $N$.  Then put links between sites $\alpha$ and $\beta$ if and only if they both have even or odd indices}.  A given triplet will contain either indices of the same parity or two of one parity and one of the other.  In the first case, there will be three links and in the other one.  This solution has $N_b$ bonds.  Proving the unicity of this construction is straightforward by induction.  First, let us assume that there exists a chain of ferromagnetic links going from sites 1 to $N-1$.  Then there must be a ferromagnetic link between the site $N$ and either site $1$ or $2$ (not both).  Both are equivalent by a permutation of models 1 and 2.  The configuration is easily shown to be determined completely and leads to the above construction.  Now, if there are no such chain between models $1$ to $N-1$, we are in one of the two cases below:
\begin{itemize}
\item There is a point with no ferromagnetic links.
\item There is a cluster of ferromagnetic links linking all models
\end{itemize}

One can check that the second of these cases leads inevitably to the existence of a chain as in the above construction and thus to the same configurations.  The first can only happen in a purely antiferromagnetic system.  This establishes the unicity of the above construction.  There are thus only two types of fixed points:
\begin{itemize}
\item Isotropic ferromagnetic fixed point (each bond is ferromagnetic)
\item  Anisotropic fixed points (arrangements of ferromagnetic and antiferrromagnetic bonds)
\end{itemize}

Having identified the different fixed points and their physical significance, we can now study the RG flows and physical relevant quantities such as critical exponents and correlation functions.
  
To simplify calculations, we considered RG flows only for the case $N=3$.  Higher dimensional behaviour is similar. We studied numerically the $g_{12}g_{13}$ projection of the flows for different values of $g_{23}$.  One can clearly view, for $g_{23}=-\gamma_3$, both the ferromagnetic and mixed points. We have presented on Figure 2, the sketch of the flow for $N=3$ in the $(g_{12},g_{13})$ plane. $F_1$ and $F_2$ are the two fixed point (plus the origin which is unstable in all the directions). A few remarks can be done.
First, note that the ``frustrated'' configurations ($g_{12}g_{13}g_{23}>0$) correspond to the half plane $g_{12}>0$ where the flow is always driven to strong couplings. Secondly, if we have the initial conditions $g_{12}^0<0,g_{13}^0>0$ and $|g_{12}|>|g_{13}|$, the flow is driven to strong coupling along the axis $g_{13}=0$. We recover a situation analogous to the case of three coupled Ising models. Namely, for $\eps={2\over 5}$ (the $3$-states Potts model), we find in the infrared limit one critical Potts model plus two  coupled Potts models which is known from integrable perturbation theory to be driven in a massive regime \cite{vays,LLM}.

\vspace{.25 cm}

\noindent {\bf Critical exponents and correlation functions}
 
In this section, we will compute the correlation functions of energy and spin operators from different models.  
To do so, we must first compute the renormalized operators $\varepsilon'$ and $\sigma'$.  In the usual way, we are looking for matrices $Z_\varepsilon$ and $Z_\sigma$ such that
\begin{equation}
\vec\varepsilon' = Z_\varepsilon \vec\epsilon
\end{equation}
\begin{equation}
\vec\sigma' = Z_\sigma \vec\sigma,
\end{equation}
where $\vec\sigma = (\sigma_1,\sigma_2,\cdots,\sigma_N)$,and similarly for $\vec\varepsilon$.  A convenient way 
to define these constants is to consider an Hamiltonian of the form
\begin{equation}
H_N = \sum_{i=1}^N H_0 - \sum_{i\ne j}^N g_{ij}\int \varepsilon_i(x)\varepsilon_j(x) d^2 x + \sum_{i=1}^N m_i \int \varepsilon_i(x) d^2 x - \sum_{i=1}^N h_i \int \sigma_i(x) d^2x,
\end{equation}
which can be seen as the original hamiltonian supplemented by source terms.  The renormalization of the source terms is then done perturbatively by expanding the coupling term.
 
Most of the involved calculations can be found in litterature \cite{DF1,DPP}, the only difference here being
 that the source terms and couplings are assumed to be different for all models and pair of models.  We will not  repeat them here.  Computing contributions and rewriting bare quantities in terms of renormalized ones, we obtain:
\begin{equation}
\frac{d \log (Z_\varepsilon(r))_{ij}}{dr} = g_{ij} - \frac{1}{2}(g_{ij})^2 + {\cal O}(g^3) 
\end{equation}
\begin{equation}
\frac{d \log (Z_\sigma(r))_{ij}}{dr} = \left(\frac{\epsilon}{8} (g^2)_{ii}\left(1+{\cal F}\right) + \frac{1}{16}(g^3)_{ii} + {\cal O}(g^4)\right)\delta_{ij},
\end{equation}      
with
\begin{equation}
{\cal F} = 2\frac{\Gamma^2(-\frac{2}{3})\Gamma^2(\frac{1}{6})}{\Gamma^2(-\frac{1}{3})\Gamma^2(-\frac{1}{6})}.
\end{equation}

We are now able to compute the correlation functions.  From the RG equations and conformal invariance of correlation functions, we have (for a lattice cut-off up to the scale $\sim R$)
\begin{equation}
\langle \varepsilon_i(0)\varepsilon_j(R) \rangle \sim \sum_{k\ne i}\sum_{l\ne j} (Z_\varepsilon)_{ik}(Z_\varepsilon)_{jl} \frac{1}{R^{2\Delta_\varepsilon^{(0)}}} \langle \varepsilon_k(0)\varepsilon_l(1)\rangle
\end{equation}
and
\begin{equation}
\langle \sigma_i(0)\sigma_j(R) \rangle \sim \sum_{k\ne i}\sum_{l\ne j} (Z_\sigma)_{ik}(Z_\sigma)_{jl} \frac{1}{R
^{2\Delta_\sigma^{(0)}}} \langle \sigma_k(0)\sigma_l(1)\rangle.
\end{equation}
If we define
\begin{equation}
(\gamma_\varepsilon)_{ij} = \frac{d \log (Z_\varepsilon)_{ij}}{dr}
\end{equation}
and $(\gamma_\sigma)_{ij}$ similarly, we can replace the renormalized quantities by their fixed point value (denoted by a star). We then get  
\begin{equation}
\langle \varepsilon_i(0)\varepsilon_j(R) \rangle \sim \sum_{k\ne i}\sum_{l\ne j} R^{-2\Delta_\varepsilon^{(0)}-(
\gamma^*_\varepsilon)_{ik}-(\gamma^*_\varepsilon)_{jl}} \langle \varepsilon_k(0)\varepsilon_l(1)\rangle
\end{equation}
and
\begin{equation}
\langle \sigma_i(0)\sigma_j(R) \rangle \sim  R^{-2\Delta_\sigma^{(0)}-(\gamma^*_\sigma)_{ii}-(\gamma^*_\sigma)_{
jj}}\langle \sigma_i(0)\sigma_j(1)\rangle.
\end{equation}
The simplifications for the spin-spin case are due to the fact that the renormalization matrix $Z_\sigma$ is diagonal.  For the energy-energy correlation functions, this is not the case and one must either diagonalize the matrix $Z_\varepsilon$ and introduce a new set of energy operators in order to define critical exponents or study long range behaviour where only the dominating term will be pertinent.  

The shift of the critical exponent $\Delta_\sigma$ is given by 
\begin{equation}
\left(\Delta_\sigma\right)_{ij} = \Delta_\sigma^{(0)} +\frac{1}{2} \left(\gamma^*_{ii}+\gamma^*_{jj}\right).
\end{equation}
Computing $\gamma^*_{ij}$ shows that the critical exponents for the anisotropic solutions are the same as the ones for the purely ferromagnetic case.  This is the case because second order contributions contain the square of the couplings and that the third order terms obey the 
\begin{equation}
g_{ij}g_{jk}g_{ki} = -\gamma_N^3
\end{equation}
condition, the same for both fixed points.  Any difference between the spin-spin critical exponent will be of order $g^4$ or higher.  Things are different for the energy-energy correlation functions, and one finds different eigenvectors and eigenvalues in the purely ferromagnetic and mixed cases.  The physical interpretation of the eigensystem is not obvious.

\vskip 0.6cm
As a conclusion, we give a brief summary of the results we found.
By using perturbative conformal field theory techniques, we considered coupled minimal models in what we believe is the most general approach.  This led us to the identification of new fixed points, which were shown, using graph theory arguments, to be, aside from the purely ferromagnetic solution,  the only non-frustrated configurations.  A study of the renormalization group flows has shown that these new points are tricritical.  Physically significant quantities such as critical exponents and correlation functions were computed perturbatively
 for the critical theories associated to the different fixed points.  Spin-spin correlation functions were shown to be identical, up to $g^4$, for the ferromagnetic and mixed configurations. Energy-energy correlation functions however, turn out to be different and can be used to distinguish between both critical behaviours.  It would be interesting to gain some numerical evidence of the existence of such fixed points for coupled 3-states Potts models. However, the tricritical nature of the mixed configurations and the fact that couplings are asymetrical and therefore that the critical line is not self-dual certainly complicate computations. 

\vspace{1 cm}
   
\noindent{\bf Acknowledgements} 

We would like to thank Vl.S. Dotsenko and G. Mussardo for helpful comments and suggestions.  M-A.L. acknowledges financial support from the NSERC Canada Scholarship Program and the Celanese Foundation.
 

\vskip 1. truecm

\noindent \underline{\bf CAPTIONS}
\vskip 1. truecm
\noindent {\bf Figure 1}: Three coupled minimal models represented by the big dots.
\vskip 0.5cm
\noindent {\bf Figure 2}: A sketch of the projected renormalisation group flow for three coupled minimal models.

\eject

\begin{figure}
\psfig{figure=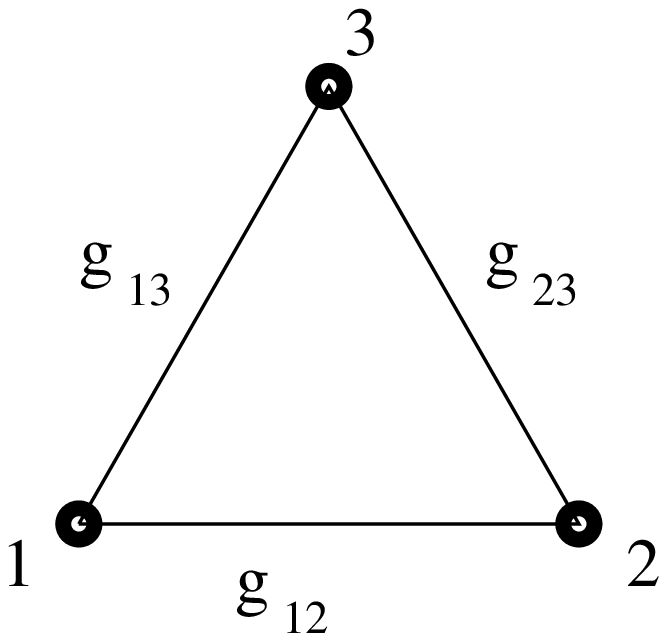,height=5cm,width=5cm}
{\bf Figure 1}
\end{figure}
\vskip 1. truecm
\begin{figure}
\psfig{figure=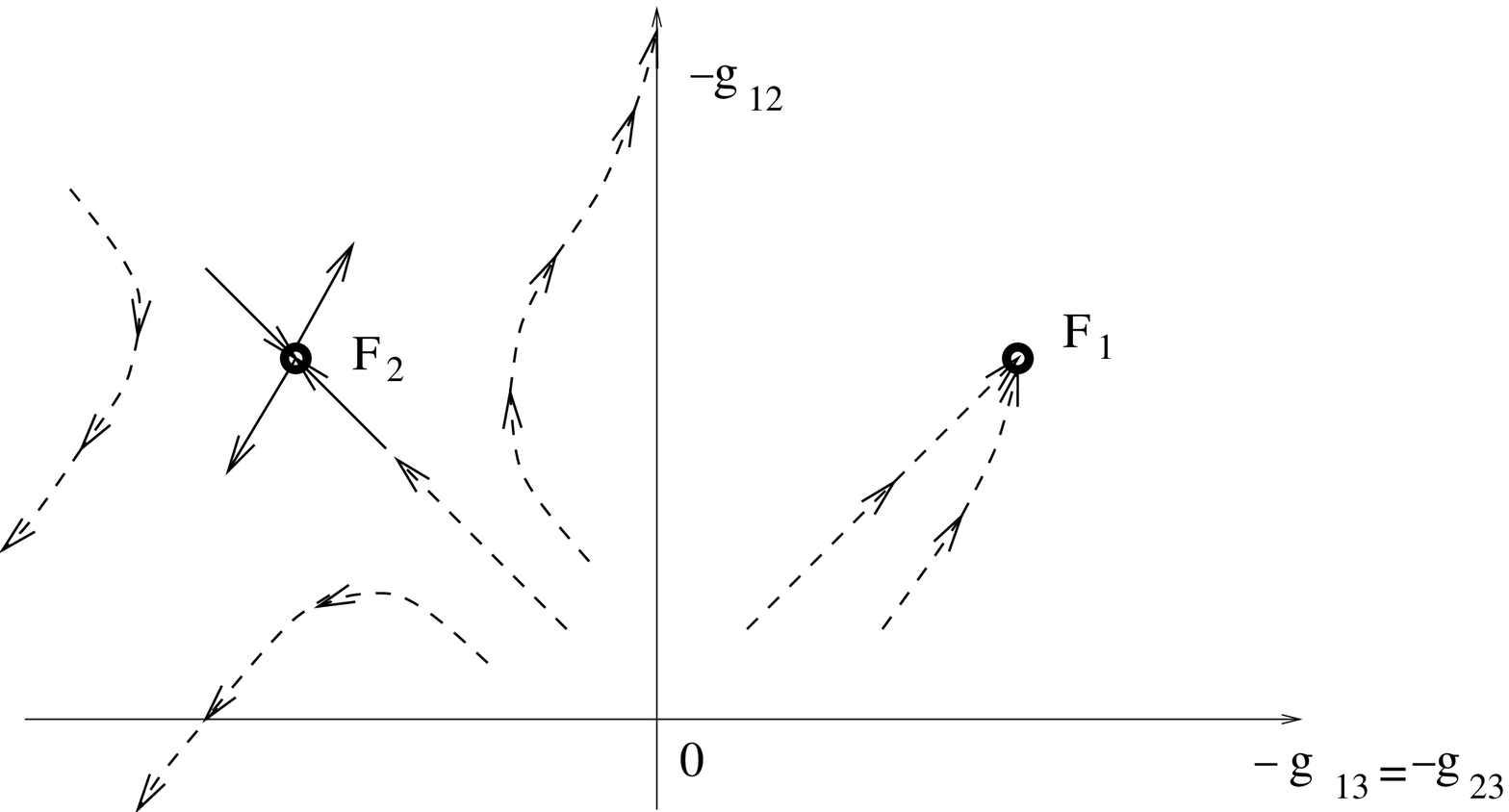,height=10cm,width=14cm}
{\bf Figure 2}
\end{figure}
\end{document}